\documentclass[12pt,preprint]{aastex}

\newcommand{\sn}{SN~2004S}
\newcommand{\kms}{km~s$^{-1}$}

\begin{document}

\shorttitle{SN~2004S Spectropolarimetry}
\shortauthors{Chornock \& Filippenko}

\title{Deviations From Axisymmetry Revealed by Line Polarization
  in the Normal Type Ia SN 2004S} 

\author{Ryan Chornock \& Alexei V. Filippenko}

\affil{Department of Astronomy, University of California,
                 Berkeley, CA 94720-3411}
\email{chornock@astro.berkeley.edu}

\begin{abstract}
We present a single epoch of high signal-to-noise ratio
spectropolarimetry of the Type Ia supernova (SN Ia) 2004S taken nine
days after maximum light.  The flux spectrum is normal, but with the
additional presence of high-velocity (HV) line features in both
\ion{Ca}{2} and \ion{Fe}{2}.  The object shows continuum polarization
at the 0.4\% level in the red, a value which appears to be typical of
SNe Ia.  The continuum data are consistent with a $\sim$10\% global
asphericity in an axisymmetric geometry.  Unlike previous observations 
of other SNe Ia with HV features, the HV features in SN~2004S show no
strong polarimetric signature, though this may be due to
the timing of our observations.  Instead, the object shows line
polarization features ($P\leq0.5$\%) that are rotated with respect to
the axis of symmetry of the continuum.  The line features 
are visible in \ion{Si}{2}, \ion{Fe}{2}, and \ion{Ca}{2}, and appear to
be narrowly confined in velocity space just above the photosphere.
These polarization features are a result of compositional
inhomogeneities in the ejecta.  They may represent newly synthesized
elements whose clumpy spatial distribution within the ejecta is
distinct from that of the globally aspherical ejecta as a whole.
\end{abstract}


\section{Introduction}

Carbon-oxygen white dwarfs that accrete enough mass to approach the
Chandrasekhar limit can explode as Type Ia supernovae (SNe~Ia).  The
first successful models of the explosion process invoked
parameterized one-dimensional (1D) flame fronts propagating outward from
the center of the star as a subsonic deflagration \citep{nom84}. 
However, the explosion process is fundamentally three-dimensional (3D)
in nature as the propagation of a deflagration front is marked by
large-scale fluid motions and mixing due to Rayleigh-Taylor
instabilities.  Recent 3D pure deflagration models are not as
successful at reproducing the 
observations because the simulated explosions are underluminous,
underenergetic, inefficient in burning the available fuel, and result
in ejecta that are too well-mixed \citep{rhn02,gam03,koz05}. 

Some modelers interpret these results as requiring the
burning front to make a transition from a subsonic deflagration to
a supersonic detonation as it becomes turbulent \citep{kh91}, though
the details of how this might happen are unclear \citep{hn00}.
Simulations, such as those of \citet{gam05}, demonstrate that models with
different prescriptions for the evolution of the burning front 
produce quite different ejecta structures.  Observational probes
of the geometry and structure of the ejecta are therefore necessary in
light of the lack of an \emph{a priori} theoretical understanding of
the explosion.

The deviations from spherical symmetry seen in the simulations
discussed above are a result of the ignition process and the random
physics of turbulent flame propagation.  More systematic departures
from sphericity could be a signature of the progenitor systems, which
are largely unknown \citep{br95,liv01,nom03}, or their environments.
Rotational flattening of the progenitor white dwarf, interaction with
circumstellar material, or the process of merging two white dwarfs
could lead to SNe~Ia with large-scale asphericities \citep{ho01,wa03}.

Spectropolarimetry provides a unique opportunity to probe the geometry
of SNe~Ia.  Outward streaming radiation in a supernova atmosphere
acquires a polarization when it is scattered by electrons.  Distant
supernovae are unresolved by the observer, so the net polarization of
the received light represents an integration of emitted flux over the
visible surface of the object.  In the case of a spherical supernova,
symmetry demands that the total polarization be zero.  Non-zero 
net polarization is therefore a sign of some deviation from spherical
symmetry in the system \citep{ss82}.

Global asphericities of the photosphere, such as ellipsoidal
deformations, are manifested as net continuum polarization
\citep{je91,ho91}.  
Past investigations of SNe~Ia have found typical intrinsic
polarizations of order a few tenths of a percent in the continuum.
The object with the highest measured intrinsic continuum polarization
percentage is still the subluminous SN~1999by \citep{ho01}, which
showed at least 0.7\% polarization.  \citet{ho01} found that the
spectropolarimetric data could be modeled by an oblate spheroid with an
axial length ratio of 1.17, seen nearly equator-on.  The lower continuum
polarizations measured in the half-dozen other SNe~Ia with good data
are consistent with global asphericities at the $\sim$10\% level
\citep{wa03,wa06,leo05,me06}.

Line polarization allows other approaches to the study of asphericity.
Polarization modulations and angle rotations across a spectral feature
could be a sign of an aspherical distribution of that
ionic species.  \citet{leo05} found that the angle rotations seen in
spectral features of some SNe~Ia could be due to clumps or ionization
asymmetries in the
ejecta.  Even if the ejecta lack such compositional inhomogeneities,
line polarization can give insight into the global asphericity of the
ejecta.  Photons scattered in a line are generally depolarized
\citep{ho96}, but in a purely ellipsoidal geometry we should expect
strong, isolated lines to be associated with ``inverted'' P-Cygni
polarization line 
profiles \citep{jef89}.  The effect of many overlapping lines is to
produce a decrease in the polarization at the same angle as the 
continuum, as was seen in SN~1999by \citep{ho01}.  These signatures of
asphericity are largely
independent of the uncertainties in the removal of interstellar
polarization (ISP) that plague estimates of the continuum
polarization. 

One exciting recent development in SN~Ia research is the association
of ``high-velocity'' (HV) features in the flux spectra with strong
spectropolarimetric features.  These HV features are so-named because
they are signatures of a separate line-formation region at higher
velocities than the normal photospheric component.  Separate HV
material was first noticed in the \ion{Fe}{2} lines of SN~1994D by
\citet{ha99}, 
but it may be ubiquitous in SNe~Ia.  Every object in the 
sample of \citet{maz05b} showed evidence of HV absorption in the
\ion{Ca}{2} near-infrared (NIR) triplet if the spectra were taken at 
sufficiently  early epochs. 

The HV \ion{Ca}{2} NIR triplet feature in the
otherwise-normal SN~2001el was responsible for a 0.7\% 
spectropolarimetric feature that was rotated with respect to the
continuum \citep{wa03,ka03}.  This feature was modeled as forming in a
spatially separate region that did not share an axis of symmetry with
the continuum \citep{ka03}.  Subsequent observations of the SNe~Ia
2004dt and 2002bf \citep{leo05,wa06} also showed strong polarization
modulations across the HV lines.  The composition and origin of this
HV material remain mysterious.  Suggestions in the literature have
included clumps of material newly synthesized in the explosion
\citep{wa03}, signatures of a gravitationally confined detonation
\citep{kp05}, and interaction with circumstellar material
\citep{wa03,ger04,maz05a}.  In all 
cases, viable models for the origin of the HV material must recognize
its intrinsically 3D nature, as revealed by the
spectropolarimetric line features and angle rotations.

Inverting the observed polarization data for one object to
construct a model is highly 
underconstrained.  A more fruitful approach is to collect data for a
sample of objects to see which, if any, model can best match the range
of observed polarimetric behaviors.  Though observations are still
sparse, \citet{leo05} have shown that SNe~Ia exhibit diverse
polarimetric characteristics that may correlate with other photometric
and spectroscopic properties.  For example, the two SNe~Ia with the
highest observed intrinsic continuum polarizations are both
subluminous, and the objects with HV features have shown the largest
line polarizations.  In addition, \citet{wa07} have claimed to find a
correlation between the strength of the spectropolarimetric modulation
across the \ion{Si}{2} $\lambda$6355 feature and the light-curve shape
(a luminosity indicator) for a sample of SNe~Ia.

In this paper we add to the small sample of SNe~Ia observed
spectropolarimetrically and present a single epoch of high
signal-to-noise-ratio (S/N) observations of \sn, a SN Ia with HV
features that is a virtual twin of SN~2001el.  The data were taken
9~d after maximum light and the observed continuum 
polarization provides evidence for asphericity in the ejecta.
Deviations from axisymmetry, seen as spectropolarimetric
modulations across prominent spectral lines, are the main result and
focus of the discussion below.

\section{Observations}

SN~2004S in MCG $-$05--16--021 was discovered by R. Martin on 2004
February 3.542 (UT dates 
are used throughout this paper; Martin \& Briggs 2004) and
spectroscopically identified as a SN~Ia by \citet{su04}.  Optical
photometry of SN~2004S has been published by both \citet{mi05} and
\citet{kk06}, with the latter authors adding a significant
near-IR light curve.  They both find \sn\ to be
photometrically normal and we adopt a date for  $B$-band maximum light
of February 4.37 $\pm$ 0.25 \citep{kk06}.
\citet{kk06} emphasize that \sn\ and SN~2001el were found to be
photometric twins.  In addition, the spectra were very similar, with
both objects showing a strong HV component to the \ion{Ca}{2} NIR
triplet lines at early times.

We observed \sn\ on 2004 February 13.34 (+9 d after maximum light)
for a total of 3600 s using the polarimeter unit\footnote{See
\url{http://alamoana.keck.hawaii.edu/inst/lris/polarimeter/manual/pol\_v3.ps}
for the online polarimeter manual by M. Cohen (2005).} of the Low
Resolution Imaging Spectrometer (LRIS; Oke et al. 1995) on the Keck I
10-m telescope.  The D560 dichroic beamsplitter sent blue light to a
400/3400 grism and transmitted red light to the 400/8500 
grating.  The 1.5\arcsec-wide slit gave a spectral resolution of about
9 \AA\ over the full observed wavelength range of 3040--9250 \AA.  The
slit was aligned at 180\degr, which was near the parallactic angle for 
the first observation.  Conditions were good and the seeing was 
$\sim$0.9\arcsec\ during these observations.

Standard two-dimensional (2D) image-processing tasks were performed using
the packages in IRAF\footnote{IRAF is distributed by the National
  Optical Astronomy 
  Observatories (NOAO), which is operated by the Association of Universities 
  for Research in Astronomy, Inc., under cooperative agreement with
  the National Science Foundation.}, as were the optimal extraction of
spectra and wavelength calibration.  We used our own routines for flux
calibration and atmospheric absorption correction of the
1D spectra \citep{ma00}.  The high airmass ($\sim$1.6)
and fixed position angle of the slit combined to make the effects of
atmospheric dispersion \citep{fi82} apparent in our
data.  As the first observation was taken close to the parallactic
angle, we adjusted the continuum slope of the combined flux spectrum to
that of the first observation by dividing the total spectrum by a
low-order spline fit to the ratio of the two spectra.

The spectropolarimetric reductions followed the standard procedure
outlined by \citet{mrg88} and implemented by
\citet{le01}.  The zero point of the waveplate was determined by
setting the observed position angle of the polarization standard star
BD +59~389 to the catalogued value of 98.09\degr\ \citep{sch92}.  As a
check on the polarization angle, we also observed a lower-quality
polarization standard, HD~127769.  The literature measurements for the
polarization vary somewhat, and the published angles are 54.0\degr\ and
52.7\degr\ \citep{ct90,mat70}.  We measured an angle of 52.5\degr,
which allows us to conclude that our observation of BD +59~389 gave a
reliable zero point.  Two null polarization standards, HD 109055 and HD
57702 \citep{berd95,ct90,mat70}, were also observed and had negligible
polarization ($P<0.07\%$). 

Dichroic absorption by interstellar dust along the line of sight
polarizes light from the supernova.  We selected two stars, CD $-31$
3659 and HD 48649, to act as probes of 
the Galactic contribution to the total ISP and observed them
immediately after the supernova.  Both are A-type stars more than 150 pc
above the Galactic plane and within 0.6\degr\ of SN 2004S; they can be regarded
as intrinsically unpolarized probes that are distant enough to sample
almost all of the dust along that line of sight out of our Galaxy,
following the suggestion of \citet{tr95}. 

The same dust that polarizes also reddens the light,
which allows us to anticipate the magnitude of the effect of ISP.  The
Galactic extinction along the line of sight has 
$E(\bv) = 0.101$ mag \citep{sfd98}.  Empirically, dust in our Galaxy
shows a polarization efficiency ($P_{MAX}$ / $E(\bv)$) of typically
3\% mag$^{-1}$, with a maximum value of 9\% mag$^{-1}$ (Serkowski et
al. 1975), so we should expect the polarization to be $\sim$0.3\%
if the line of sight is typical, up to a maximum of 0.9\%.  We
integrated the flux-weighted Stokes parameters over 5050--5950 \AA\ to
approximate the $V$-band polarization ($q_V$,$u_V$) and measured
(0.03, 0.49) and (0.11, 0.30), respectively, for the two probe stars.
The two stars have somewhat different polarization, so we averaged
their Stokes parameters to form our estimate of the Galactic
ISP.  Smoothed versions of $q$ and $u$ were then subtracted from 
the \sn\ observations.  Uncertainties in the determination of the
Galactic ISP are absorbed into the determination of the host-galaxy
contribution to ISP below.

\section{Results}

\subsection{Total-Flux Spectrum}\label{fluxsect}
The high S/N total-flux spectrum is presented in
Figure~\ref{compfig} along with a spectrum of the prototypical normal
Type Ia SN~1989B \citep{we94}.  We have removed a radial velocity for
the host galaxy of \sn\ of 2808 $\pm$ 5 \kms\ (determined from 21-cm
observations by \citet{th05}) from all plots in this paper.  The SN 
1989B spectrum is from 8~d after $B$-band maximum light and is very
similar to our \sn\ spectrum, confirming that the age inferred from
the photometry is accurate.  \citet{we94} measured the photometric
decline-rate parameter \citep{ph93} for SN~1989B to be $\Delta
m_{15}(B)=1.31 \pm 0.07$ mag, compared to the value of $1.14 \pm 0.01$
mag reported by \citet{kk06} for \sn. 

We shall use the fiducial SN~1989B spectrum as a reference to better
understand \sn.  The
most obvious difference between the two spectra is that the
absorptions due to \ion{Ca}{2} are significantly deeper in \sn\ and
extend to much higher velocities.  \sn\ otherwise appears to be
spectroscopically quite normal for a SN~Ia \citep{fil97}.  A more
complete spectroscopic sequence is presented in \citet{kk06}, along
with comparisons to a suite of SNe~Ia.

Figure~\ref{linefig} shows a closer view of the spectral features due
to \ion{Ca}{2} and \ion{Si}{2}.  The \ion{Si}{2} $\lambda$6355 line in
particular is the hallmark of SNe~Ia and is frequently used to measure
the photospheric velocity (e.g., Branch et al. 1988).
As shown in the middle panel of Figure~\ref{linefig}, the absorption
minimum of the \ion{Si}{2} line in SN~1989B 
is at $\sim$10,000 \kms, a typical value for a SN~Ia.  The
\ion{Si}{2} feature in \sn\ is noticeably weaker and the minimum is at
only 8900 \kms, compared to the figure of 9300 \kms\ reported by
\citet{su04} near $B$-band maximum light.  This velocity is 
actually quite low for a normal SN~Ia at this epoch and is comparable
to the lowest-velocity objects in the sample of \citet{ben05}.  Only
peculiar subluminous objects such as SNe~1991bg and 2002cx have
shown lower \ion{Si}{2} velocities at a similar age \citep{lei93,li03}.

By contrast, Figure~\ref{linefig} shows that the \ion{Ca}{2} spectral
features in \sn\ were formed at higher velocities than in SN~1989B.
The \ion{Ca}{2} H\&K $\lambda\lambda$3934, 3968 absorption feature
in SN~1989B has a minimum at 
3795~\AA\ (11,600 \kms) and a rather flat-bottomed absorption with the
blue wing rising to a flux 
maximum at 3675~\AA\ (21,200 \kms), where the velocities
reported here are relative to the $gf$-weighted line wavelength of
3945~\AA.  In \sn, the absorption is much deeper and the flux minimum
is much bluer, at 3710 \AA\ (18,400 \kms), with absorption
extending outward to 3600 \AA\ (27,400 \kms). 

The profile of the \ion{Ca}{2} NIR triplet shown in
Figure~\ref{linefig} is more complicated than the other two lines,
with at least three distinct flux minima present near 8080, 8230, and
8385 \AA.  Two effects contribute to the complexity of this line
profile.  The first is that the feature is intrinsically formed by a
triplet and the wavelength separation of the components is relatively
large.  The minima at 8385 and 8230 \AA\ can then be identified with
\ion{Ca}{2} $\lambda$8662 and $\lambda\lambda$8498, 8542,
respectively, with a blueshift of 10,000--11,000 \kms.  Absorption minima
at similar velocities can be be seen in the 
reference spectrum of SN~1989B.  \sn\ shows an additional flux minimum
near 18,000 \kms\ with no corresponding feature in SN~1989B, which we
identify as the counterpart of the 
high-velocity absorption seen in \ion{Ca}{2} H\&K.  There are 
therefore two \ion{Ca}{2} line-forming regions in \sn, one at the
photosphere and a second one separated from the photosphere by 8,000
\kms.  No photospheric component is visible in the H\&K feature, which
is dominated by the HV component.  The spectral sequence of
\sn\ presented by \citet{kk06} shows the evolution of the \ion{Ca}{2}
NIR triplet absorption profile from being dominated by HV material at
early times to photospheric velocities at late times.

The exact velocity
range of this HV absorption is difficult to determine directly from
the spectrum because of confusion caused by line blending.  If the
minimum at 8080 \AA\ is identified with $\lambda$8542, then the
apparent velocity is 16,700 \kms, but at that velocity the
$\lambda$8662 line would produce an absorption at 8190~\AA, which is
close a local maximum in the line profile.  It is likely that the HV
material is at slightly higher velocity (such as 18,000 \kms), but with
a velocity spread sufficient to blend the two components of the line
profile together.  A spectral synthesis calculation is necessary to
make a more quantitative estimate of the velocity spread of the HV
material.  The maximum velocity of the 
absorbing material is easier to isolate, as the blue edge of the
feature is at 7880~\AA\ (24,000 \kms\ if identified with $\lambda$8542).

As with all supernova spectral features, blends with other lines due
to the large Doppler shifts may
affect the strength and apparent wavelength of any given spectral
feature.  In particular, \ion{Si}{2} $\lambda$3858 and \ion{O}{1}
$\lambda$8446 are candidates for making a potential contribution to
the H\&K and NIR triplet features, respectively.  We do feel confident
in identifying the presence of a high-velocity absorption component in
\ion{Ca}{2}, especially with the simultaneous detection of a feature
at a similar velocity in both the H\&K and NIR triplet features.  We
note that no HV component is obvious in the profile of the \ion{Si}{2}
$\lambda$6355 line.  There are notches in the blue wing, but these are
the early signs of incipient iron lines that strengthen with time, not
HV features (see the spectral sequence in Krisciunas et al. 2007).

\subsection{Spectropolarimetry}

The last step of preparation before we can begin analysis of the
spectropolarimetry is to correct the data for ISP in the host galaxy.
Normal SNe~Ia have well-defined colors as a function of age, so
\citet{mi05} were able to estimate a total $E(\bv)$ to \sn\ of 0.18
$\pm$ 0.05 mag by comparison with templates.  The numbers given above
imply that the host galaxy 
contributes $E(\bv) \approx 0.08$ mag of reddening, so we
expect a typical host ISP of $\sim$0.24\% \citep{ser75}.  \citet{kk06}
used the long baseline of their optical through NIR observations
to set a tighter 
constraint on the extinction.  They derive a host extinction of $A_V =
0.071 \pm 0.043$ mag, which would imply a host ISP of 0.07\% for
typical dust, up to a maximum of 0.21\%.  Unlike the Galactic
ISP, we have no method to directly measure the host contribution, so
we will have to appeal to some interpretation of the SN polarimetry
itself to make an ISP estimate.  This ISP estimate will not be unique
because different assumptions about the intrinsic SN polarization can
lead to different ISP estimates.

Some of the \sn\ spectropolarimetry data are plotted in the Stokes $q-u$
plane in Figure~\ref{qufig}.  Each circle represents a wavelength bin
50 observed \AA\ wide and the size of each is proportional to the
wavelength of the bin it represents, so larger 
circles are from longer wavelengths.  Anticipating the results below,
we only plot data points from sections of the spectrum that we believe
are free from strong line polarization features.  The exact wavelength
choices for these ``continuum'' data points are marked in the middle
panel of Figure~\ref{spolfig}.  The wavelength dependence of ISP is
sufficiently weak at optical wavelengths \citep{ser75} that the effect
of subtracting a small ISP is similar to choosing a new origin for the
data in the $q-u$ plane.  The distribution of plotted wavelength  
bins is elongated, with a clear spread of at least 0.4\% in polarization
between the blue end of the spectrum (smaller points) and the red end
(larger points), which must be due to intrinsic SN polarization.  The
dotted circle in Figure~\ref{qufig} has a radius of 0.21\%,
corresponding to the maximum expected host ISP. 

The line plotted in Figure~\ref{qufig} is a fit to the data points.
The simplest decomposition of the polarization is to choose an ISP
vector that lies on the line and to assign the remaining
polarization to intrinsic SN polarization that varies in amplitude as
a function of wavelength,
with most of the variation occurring along the axis defined by the
line.  The direction indicated by the line could then represent the 
symmetry axis of the SN continuum as projected onto the sky
\citep{wa01}.  This argument does not give a unique value for the ISP,
so we will appeal to SN~Ia theory and spectral modeling calculations to
assume that the intrinsic polarization at blue wavelengths must be
small.  As \citet{ho01} have argued, many overlapping lines of
iron-peak elements are the dominant form of opacity in SNe~Ia at 
wavelengths below 6000 \AA.  The opacity from these lines should
destroy any polarization caused by electron scattering in an
aspherical SN atmosphere.  \citet{ho01} then chose an ISP vector to
make the polarization zero at the blue end of their data.  Other
authors \citep{leo05, wa06, me06} have chosen ISP vectors to make the
intrinsic polarization of certain blue emission features zero.

For this work, we
will select an ISP vector ($q_{ISP}$,$u_{ISP}$) that lies on the line
in Figure~\ref{qufig}.  The exact point along this line is arbitrary,
but we chose the point ($-0.040$, $-0.159$), which we have marked with a
square, in order to make the polarization small at blue wavelengths.
A host ISP of this magnitude is consistent with the reddening
constraints mentioned above.  We subtracted an ISP with the
wavelength dependence of \citet{ser75} and assumed $R_V = A_V/E(B-V) 
=3.1$.

The formal polarization [$P = (q^2 + u^2)^{\frac{1}{2}}$] is
positive-definite and is therefore biased high for low values of the
polarization.  We can avoid this bias by continuing to represent the
data by two Stokes parameters, but we are free to choose a more
convenient parameterization of rotated Stokes parameters (RSP;
Trammell et al. 1993).  In this case, we will
rotate the ISP-subtracted data to align one Stokes parameter
(called $q_{RSP}$) with the obvious axis seen in the continuum data in
Figure~\ref{qufig} and marked by the line.  This coordinate system is
similar to the decomposition of polarization into a dominant and
orthogonal axis by \citet{wa03}, except that we have explicitly
excluded wavelengths of strong line polarization from the fit.

The rotated \sn\ spectropolarimetry is plotted in
Figure~\ref{spolfig}.  After the rotation, most of the continuum
polarization is in the $q_{RSP}$ Stokes parameter.  The polarization
rises from near zero at 4000 \AA\ to a maximum of 0.4\% at 6800 \AA\
before declining at longer wavelengths.  Several broad undulations in
the continuum polarization are apparent as well as narrower features 
associated with spectral lines seen in the flux spectrum.  Five of
these narrower features 
are also visible as strong negative departures from the continuum in
the $u_{RSP}$ parameter and hence as rotations in $\theta$.  Spectral
windows away from these narrower features are mostly flat and near
zero in $u_{RSP}$ despite the broad undulations in $q_{RSP}$,
justifying our decomposition of the polarization into these Stokes
parameters.  We now consider the line features in more detail.

\subsubsection{Silicon}

We will first examine the polarization of the \ion{Si}{2}
$\lambda$6355 line because it is the most isolated and can be used as
a reference to understand the other spectropolarimetric features.
As noted above, the continuum polarization of \sn\ is confined to
$q_{RSP}$, while Figure~\ref{spolfig} shows that the \ion{Si}{2} line
has excursions away from the continuum polarization in both $q_{RSP}$
and $u_{RSP}$.  We can immediately conclude that the
\ion{Si}{2} line-formation region does not share an axis of symmetry
with the continuum. 

This point is reinforced in the upper-left panel of
Figure~\ref{linepolfig}, where data points from both the continuum and
the line are plotted in the $q-u$ plane.  The continuum points are the
same as in Figure~\ref{qufig}, but rotated into the new coordinate
system and shown in gray.  The elongation of the continuum points
along a line in the $q-u$ plane was argued above to be representative
of scattering in a geometry with a single well-defined axis.  The data
points in black with error bars are 50 \AA\ bins between wavelengths
of 5900 and 6400 \AA.  These line polarization points clearly do not
share the same distribution in the $q-u$ plane as the continuum.  They
are arranged along a linear axis that is rotated with respect to, and
intersects the major axis of, the spread of continuum points.

Another view of the same data can be seen in the leftmost panels of
Figure~\ref{multifig}, where the Stokes parameters are plotted as a
function of velocity in the line in 20 \AA\ bins.  The \ion{Si}{2}
$\lambda$6355 line is located in the middle of a broad depression in
$q_{RSP}$, which has local maxima near 5700 and 6800 \AA.  There is a
narrower, sharp feature in $q_{RSP}$ located just blueward of the
minimum of the line in the flux spectrum.  This feature is at a
similar velocity as the largest negative depression in $u_{RSP}$.  The
flux minimum of the 
line is at 8900 \kms\ and the greatest deviation of the line
polarization from the continuum is in the narrow range of approximately
9,000-11,000 \kms, just above the photosphere.

Interestingly, as one moves to redder wavelengths from the flux minimum
of the $\lambda$6355 line, $u_{RSP}$ increases from its minimum,
crosses zero near 6350 \AA\ ($\sim$0 \kms), 
and rises to a peak in the red wing of the emission component of the
line.  This positive excursion in $u_{RSP}$ returns to zero (the
continuum value) near 6700 \AA, corresponding to a redshift of
$\sim$16,000 \kms, which may be a spectropolarimetric indication of
\ion{Si}{2} forming in the far side of the ejecta almost as far out as
the expansion velocity derived above for the HV line-formation region.

\subsubsection{Calcium}

The next ionic species we examine is \ion{Ca}{2}.  The analysis of the
flux spectrum above presented evidence for two \ion{Ca}{2} line-formation 
regions, at 10,000 and 18,000 \kms, making the shape of the NIR triplet
in the flux spectrum complicated.  The polarization profile of the NIR
triplet shown in Figure~\ref{spolfig} is also complex.  Two separate
polarization features are seen in both $q_{RSP}$ and $u_{RSP}$ near
8165 and 8345 \AA.

The presence of two line-formation regions makes interpretation of the
polarimetric features ambiguous, but the rightmost panels of
Figure~\ref{multifig} can clarify the situation.  The three lines
labeled ``HV'' in the top right panel mark the expected location of
features associated with the three components of the triplet,
blueshifted by the 18,000 \kms\ velocity of the HV line-formation
region seen in the total-flux spectrum.  The stronger of the two
polarimetric features is at the same velocity as the expected
$\lambda$8662 line, but no feature is seen at the expected position of
the other two lines.  A much more natural identification of the lines
can be seen in the lower-right panel.  The lines labeled ``PV'' mark
the expected location of 
features associated with the NIR triplet if they occur at the same
velocity (10,000 \kms) as the strongest polarization feature in the
\ion{Si}{2} line.  The redder of the two polarization features is
formed over a velocity range of 9,500$-$12,500 \kms\ if it is
identified as \ion{Ca}{2} $\lambda$8662.  The bluer of the two
features is then a blend of $\lambda$8498 and $\lambda$8542 forming at
similar velocities.  In this interpretation, the two polarimetric
features are formed at velocities just above the local minima in the
total-flux spectrum ($\sim$11,000 \kms), similar to the behavior of
the \ion{Si}{2} $\lambda$6355 line discussed in the previous 
section.

The data points plotted in the $q-u$ plane in the lower-left panel of
Figure~\ref{linepolfig} are 50 \AA\ bins
between 8050 and 8300 \AA\ and correspond to the polarization
feature we associated with $\lambda\lambda$8498, 8542.  Data points
from the $\lambda$8662 feature, spanning 8300 to 8500 \AA, are plotted
in the lower-right panel. In each panel the data points are connected
in order of wavelength by a line.
The location of these two features in the $q-u$ plane
reveals complex spectropolarimetric behavior.
Not only do these \ion{Ca}{2}
$\lambda\lambda$8498, 8542 data points show a spread of ~0.4\% in each
Stokes parameter, which is itself indicative of deviations from simple
axisymmetry, but the line connecting the data points shows that the
path in the $q-u$ plane is ordered in the form of a loop.  This is in
contrast with the \ion{Si}{2} data, which form a linear feature.  The
$\lambda$8662 feature shows a similar 2D spread with a
suggestion of a loop, though with lower statistical significance.

\subsubsection{Iron}

Next we will consider the two features visible in Figure~\ref{spolfig}
as deviations from the continuum in both $u_{RSP}$ and $\theta$ near
4800 \AA.  The spectra of SNe~Ia in this wavelength region are formed by
a complicated blend of lines (mostly \ion{Fe}{2}, \ion{S}{2}, and
\ion{Si}{2}), so the identity of any given spectral feature is
ambiguous.  The wavelengths of the 
strongest rotations in these features are at about 4745 and 4832 \AA.
We first assume that the spectropolarimetric features are formed in a
similar manner to the \ion{Si}{2} line analyzed above.  The narrow
core of the polarization feature in the \ion{Si}{2} line was centered
at an expansion velocity of $\sim$10,000 \kms.  De-blueshifting the
unknown features by this velocity implies rest wavelengths of 4906 and
4996 \AA\ for our mystery lines.  The \ion{Ca}{2} NIR triplet polarization
feature was formed at a velocity of 11,000 \kms, which would imply
rest wavelengths of 4922 and 5013 \AA\ for these lines.

One reasonable identification is with \ion{Fe}{2} $\lambda\lambda$4924,
5018 if those lines are blueshifted by $\sim$11,000 \kms.  There are 
two major complications for this interpretation.  The first is that
those \ion{Fe}{2} lines form a triplet with $\lambda$5169.  All three
lines share the same lower atomic level, so they should all be present
if one of them is.  With this identification, the stronger
spectropolarimetric feature is associated with $\lambda$4924, the
weaker one with $\lambda$5018, and $\lambda$5169 lacks a counterpart.
This ordering by strength of spectropolarimetric 
features is the reverse of the ordering by weighted oscillator strengths
($gf$).  Naively, one might expect the strongest line to produce the
strongest polarimetric feature, much as the strongest of the three
lines produces the strongest feature in the flux spectrum.

\begin{figure}
\plotone{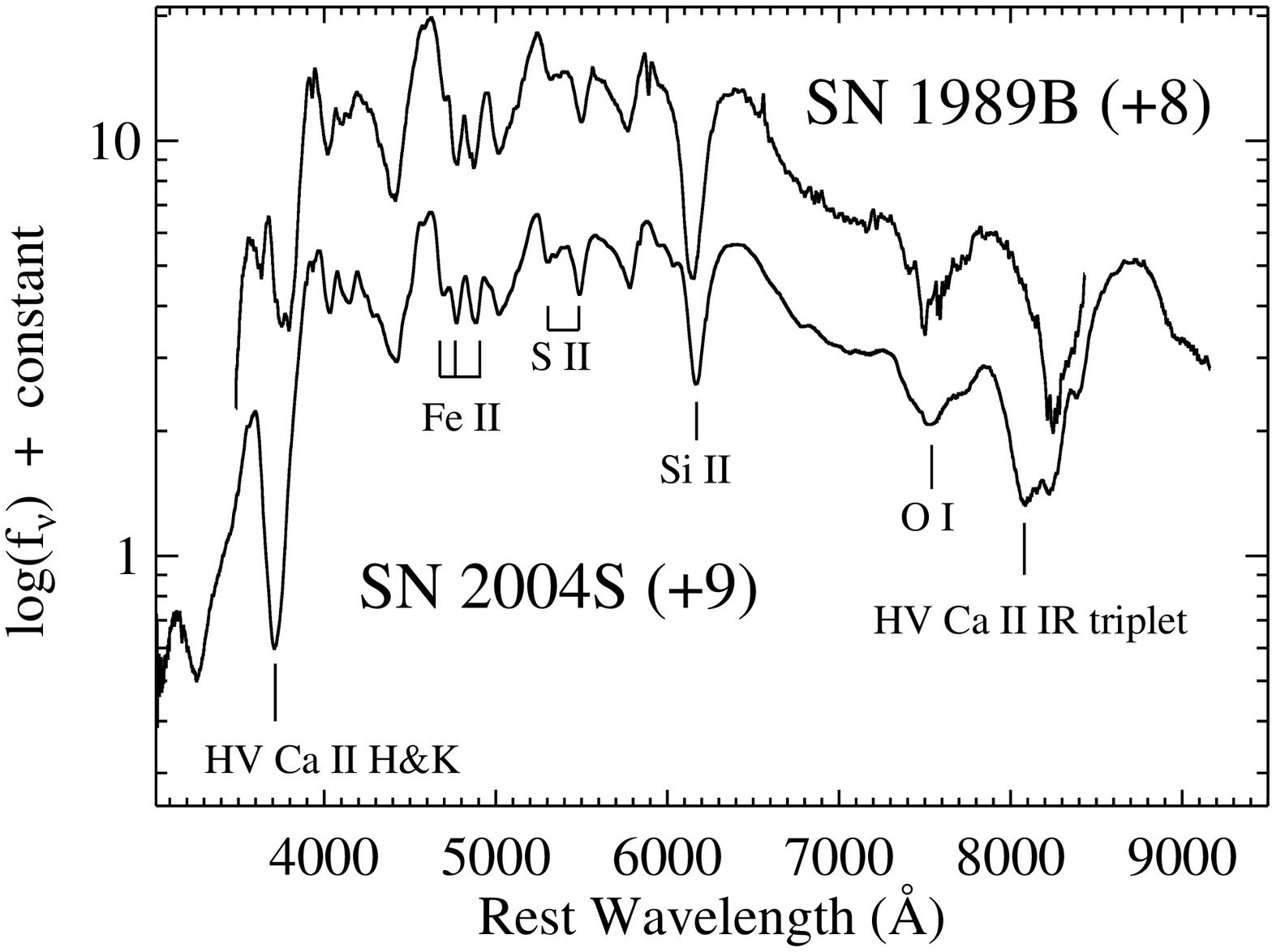}
\caption{Total-flux spectra of SN~2004S and the normal Type Ia
  SN~1989B from 1989 February 15 \citep{we94}.  The numbers in
  parentheses are ages 
  relative to $B$-band maximum light.  Some lines of polarimetric
  interest are labeled.  The major difference between the two spectra
  is that SN~2004S clearly shows an extra high-velocity (HV) absorption
  component in both of the spectral features due to \ion{Ca}{2}.  See
Figure~\ref{linefig} for details of these line features.  The
locations of the members of the \ion{Fe}{2} triplet are marked at a
blueshifted velocity of 16,000 \kms.
}
\label{compfig}
\end{figure}

\begin{figure}
\plotone{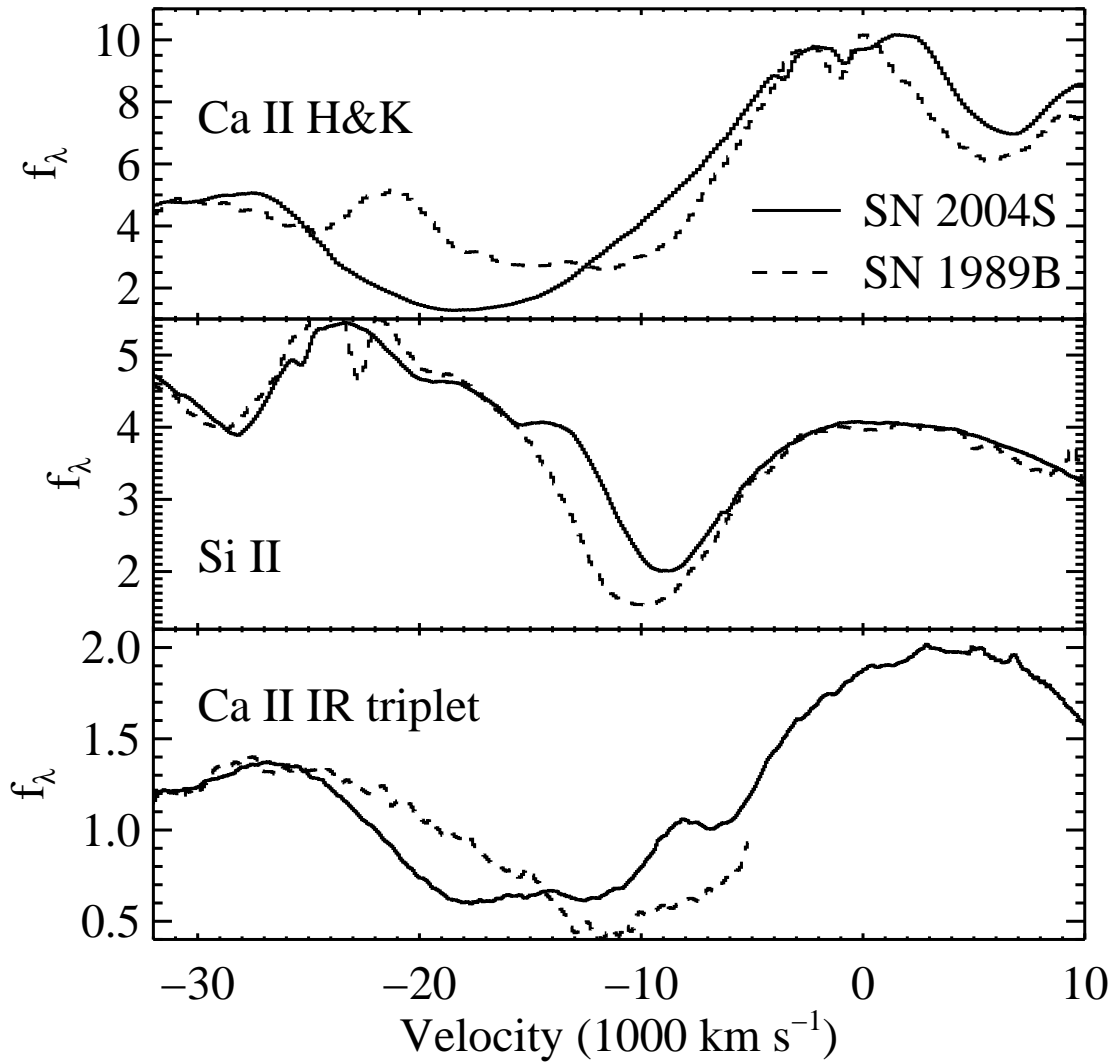}
\caption{Prominent absorption lines in SN~2004S (solid) and SN~1989B
(dashed).  The velocity axis in each panel is computed relative to the
$gf$-weighted line centroid for each blend (3945 \AA, 6355 \AA, and
8579 \AA, respectively).  A separate multiplicative factor has been
applied to the flux scale of the spectra in each panel to match
the flux peaks of the two objects in order to facilitate comparison of
the absorption features.  In SN~1989B, all three lines have flux
minima at similar velocities (10,000--12,000 \kms), while in SN~2004S
an additional absorption component is visible in both of the \ion{Ca}{2}
features at 18,000 \kms, well above the
photospheric velocity defined by the \ion{Si}{2} line.
}
\label{linefig}
\end{figure}

\begin{figure}
\epsscale{0.95}
\plotone{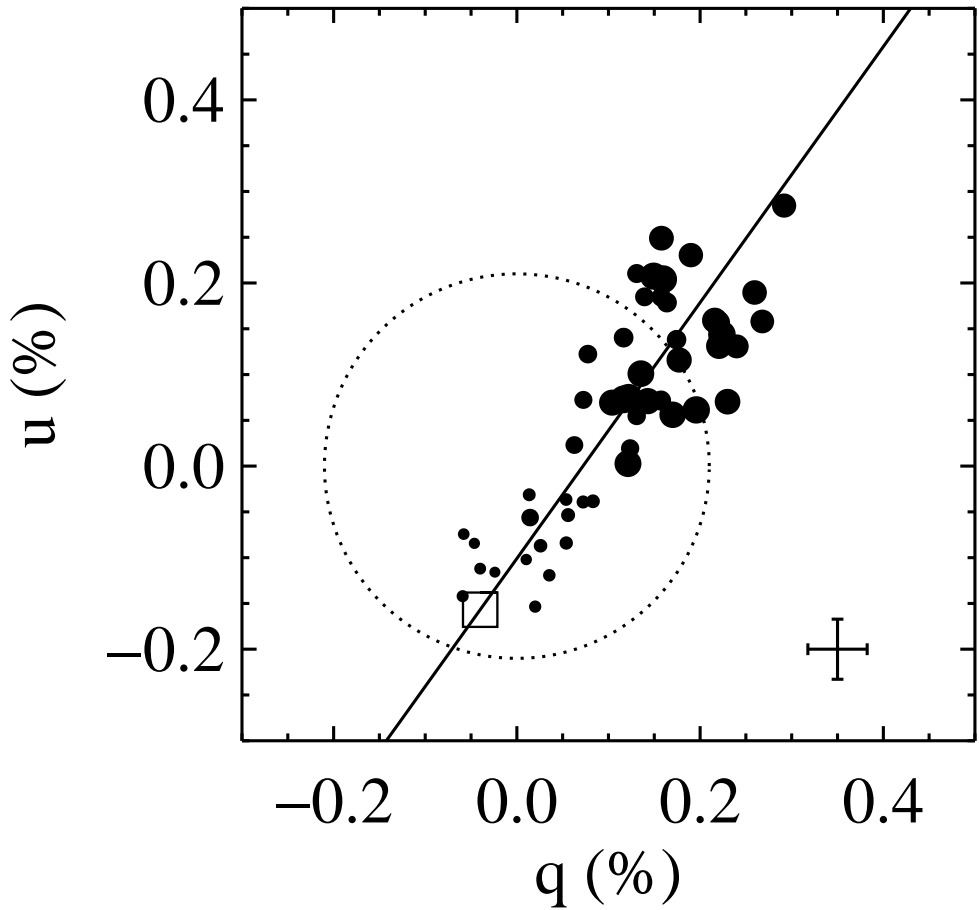}
\caption{Continuum polarization data for SN~2004S in the $q-u$ plane.
  The data have been corrected for the Galactic component of ISP, as
  determined from probe stars.  Each circle represents a bin of width
  50 observed \AA\ selected from wavelength regions free of strong
  line polarization.  See 
  Figure~\ref{spolfig} for the definition of these regions.  The size
  of the circles is proportional to wavelength, so a gradient in
  polarization is evident between the shorter and longer wavelengths.
  The distribution of points is consistent with a single dominant axis
  of symmetry for the continuum.  The line 
  is a fit to the data points and the box marks a point
  along the line chosen to represent the host-galaxy contribution to
  the ISP.  The dotted circle has a radius of the maximum expected
  ISP from host reddening arguments.  After subtracting the chosen ISP
  vector, the data were rotated to 
  align the new Stokes parameter $q_{RSP}$ with the axis of symmetry
  represented by the line.  The cross 
  at (0.35, $-$0.2) shows the median statistical error bars of the
  points in the plot.
}
\label{qufig}
\end{figure}

\begin{figure}
\epsscale{0.75}
\plotone{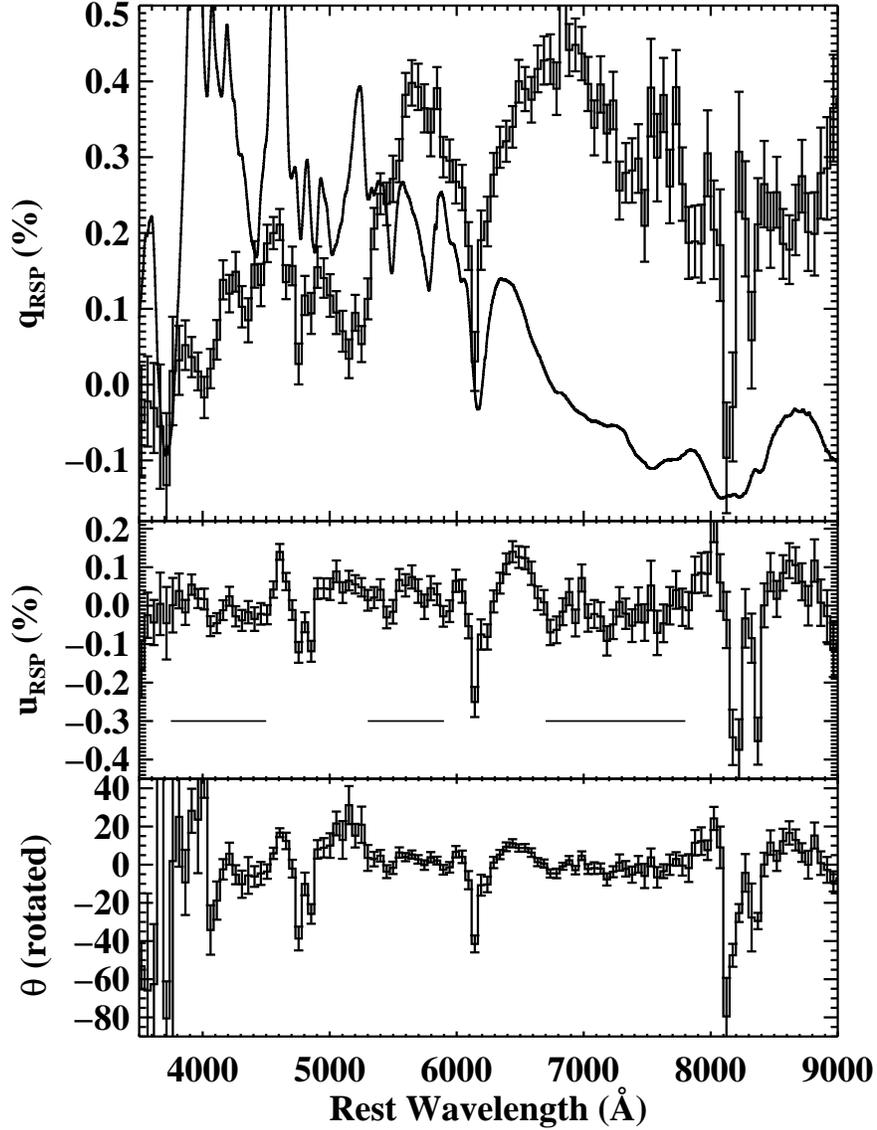}
\caption{Spectropolarimetry of SN~2004S.  The polarization data
  plotted here have been corrected for ISP and binned to 50 observed
  \AA\ per bin for display purposes.  In
  addition, the Stokes parameters have been rotated to a new
  coordinate system ($q_{RSP},u_{RSP}$), as described in the text.
  The total-flux spectrum is plotted in the background of the top panel to
  guide the eye.  Rotations due to line features of \ion{Fe}{2},
  \ion{Si}{2}, and \ion{Ca}{2} are visible in $u_{RSP}$ and $\theta$.
  The horizontal lines overplotted in the 
  middle panel represent sections of the spectrum that are free from
  strong line polarization features and are used to define
  ``continuum'' polarization data points in Figures~\ref{qufig} and
  \ref{linepolfig}.  Note that the angle is ill-defined for wavelengths
  less than 4200 \AA\ due to the polarization data points being close
  to the origin of this coordinate system.
}
\label{spolfig}
\end{figure}

\begin{figure}
\epsscale{1.0}
\plotone{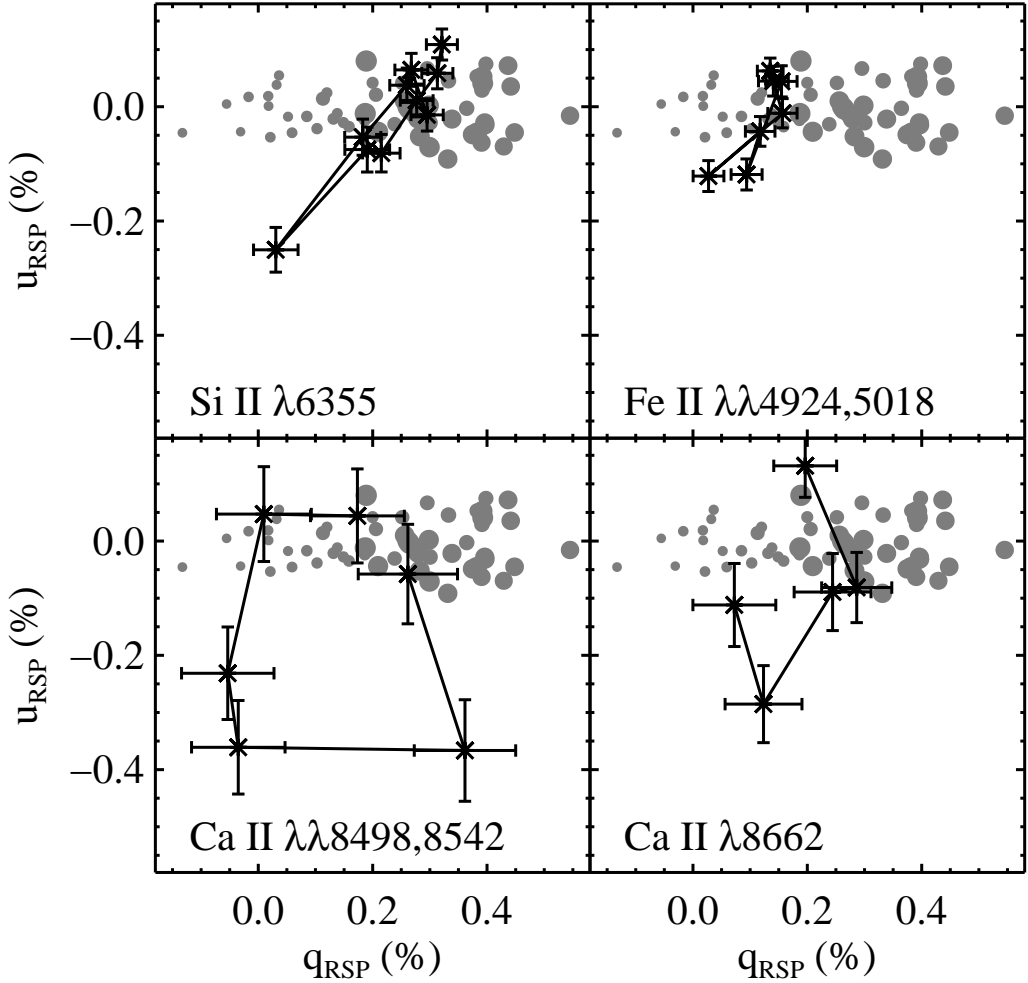}
\caption{Polarization of various spectral features in the $q-u$ plane.
  The gray circles in the background of each panel represent the same
  wavelength bins from the continuum windows marked on
  Figure~\ref{spolfig} and shown in Figure~\ref{qufig}, but rotated to
  the new coordinate system.  The continuum points are clearly
  extended along a single line in the $q-u$ plane, indicative of a
  single axis of symmetry.  The points representing wavelength bins
  associated with spectral features due to each ion (shown in black
  with error bars) fall outside the spread of continuum points,
  implying that the lines do not have the same spatial distribution
  as the continuum.  The \ion{Si}{2} and \ion{Fe}{2} features show
  similar polarization behavior that may reflect a common spatial
  origin.  The \ion{Ca}{2} data points are inconsistent with a 
  simple single-axis symmetry because they show a spread in
  both $q_{RSP}$ and $u_{RSP}$.  The feature associated with
  \ion{Ca}{2} $\lambda\lambda$8498, 8542 appears to form a
  loop in the $q-u$ plane.
}
\label{linepolfig}
\end{figure}

\begin{figure}
\epsscale{1.15}
\plotone{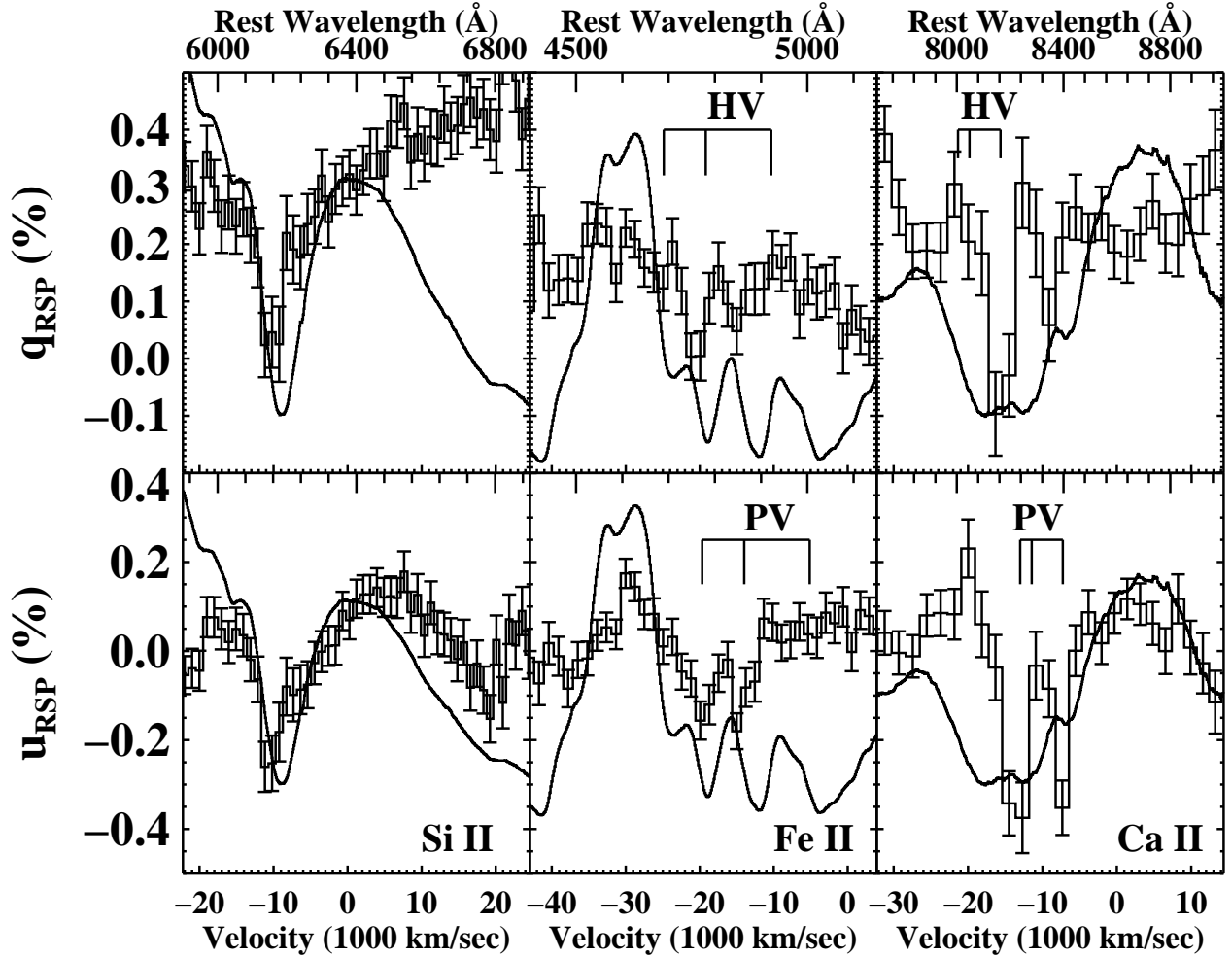}
\caption{Stokes parameters versus velocity for three species.  The
  total-flux spectrum is overplotted in each panel to guide the eye.
  The line 
  velocities used for the coordinates are relative to the $gf$-weighted
  line wavelength for each ionic blend.  \ion{Si}{2}
  $\lambda$6355 shows a broad polarization depression with a
  superposed narrow feature, present in both $q_{RSP}$ and $u_{RSP}$,
  and centered at 10,000 \kms, just blueward of the photosphere.  To
  aid in the identification of spectropolarimetric 
  features, overplotted on the \ion{Fe}{2} and \ion{Ca}{2}
  panels are the expected wavelengths of each member of the respective
  triplets, blueshifted both by the photospheric-velocity (PV) and
  the high-velocity (HV) line-formation region for each ion.  The
  polarimetric features are clearly associated with the PV for both
  ions, except for the lack of a feature associated with \ion{Fe}{2}
  $\lambda$5169.  There are no strong polarimetric features associated
  with the HV line-formation region.  See text for details.
}
\label{multifig}
\end{figure}

The second complication can be seen in the flux spectrum in
Figure~\ref{compfig}.  A series of flux minima at 4698, 4772, and 4885
\AA\ are labeled as \ion{Fe}{2}.  If this identification is correct,
the three components of the triplet are blueshifted by 14,000, 15,000,
and 17,000 \kms, respectively.  These velocities are much greater than
the photospheric velocity measured from the \ion{Si}{2} line, similar to
the HV \ion{Ca}{2} features discussed above, and thus raise the specter
of multiple \ion{Fe}{2} line-forming regions, one at photospheric
velocity and one at high velocity.  HV \ion{Fe}{2} was
first invoked by \citet{ha99} to model the early-time spectra of SN~1994D
and has since been used by several authors to explain spectra of
SNe~Ia before and near maximum light (e.g., Stehle et al. 2005).
\citet{br06} found that their spectral fits to a number of SNe~Ia were
improved with the addition of HV \ion{Fe}{2}.  Suggestively, they also
found that the object in their sample whose fit required the highest
optical depth in HV \ion{Fe}{2} was SN~2001el, which also had strong
HV \ion{Ca}{2} absorption.  Spectral fits that lack HV \ion{Fe}{2}
have difficulty producing alternative identifications for these
absorption minima in SNe~Ia.

These difficulties with the \ion{Fe}{2} interpretation lead to the
natural question of whether any other ionic species could be
responsible.  The first alternative identifications we consider
are \ion{Si}{2} and \ion{Ca}{2} because these species are known to
produce polarized spectral 
features at other wavelengths in this object.  \ion{Ca}{2} has no
strong spectral features in this wavelength region, but the 
\ion{Si}{2} $\lambda$5051 blend is a 
plausible candidate for creating a spectropolarimetric feature.
\citet{wa06} identified a feature from this line
in SN~2004dt, but in \sn\ this single line would have to explain two
polarimetric features.  These features would be at expansion
velocities of 18,700 and 13,300 \kms, significantly different from the
polarimetric behavior of the strong \ion{Si}{2} line ($\lambda$6355),
so we reject the \ion{Si}{2} interpretation.

The most plausible remaining candidate ion is \ion{S}{2}, which has a
large number of transitions in the optical.  Blends of these lines,
particularly between 4500 and 5700 \AA, make a contribution to the
spectra of SNe~Ia.  The strongest spectroscopic
feature in the flux spectrum attributed to \ion{S}{2} in SNe Ia is the
``W''-shaped absorption near 5300 \AA\ (see Figure~\ref{compfig}).  No
strong spectropolarimetric feature is visible in 
Figure~\ref{spolfig} near this wavelength.  An identification of
the polarimetric features with \ion{S}{2} would have to explain why
the weaker lines leave a polarimetric signature and the stronger ones
do not.

We will adopt the identification of these lines with \ion{Fe}{2} for
the rest of the paper, noting the caveats above.  Data points from
these two lines (50 \AA\ bins between 4650 and 5000 \AA) are plotted
in the upper-right panel of Figure~\ref{linepolfig}.  The distribution
of points in the $q-u$ plane is elongated in a linear manner along a
direction roughly parallel to the linear feature seen in \ion{Si}{2},
possibly indicating a common origin for the spectropolarimetric
features.  The middle panels of Figure~\ref{multifig} show the same
data in velocity space.  The expected locations of the three members of
the triplet if they were present at the 15,000 \kms\ velocity inferred
from the flux spectrum are marked in the $q_{RSP}$ panel, while the
$u_{RSP}$ panel shows the expected position of features at 10,000
\kms\ as measured from the \ion{Si}{2} polarization feature.  The
polarimetric line features do not fall at the same position as the
HV notches in the flux spectrum, but they do correspond to the
positions of two of the three lines at photospheric velocities.

One possible
explanation for the lack of a polarimetric counterpart to $\lambda$5169
is to rely on the observed narrowness of the sharp 
spectropolarimetric feature seen in the core of the \ion{Si}{2} line
and at the photospheric velocity in the \ion{Ca}{2} NIR triplet.
Perhaps whatever non-axisymmetric asphericity (e.g., clumping) is the
cause of the spectropolarimetric line features and rotations in the
other species is narrowly confined in the ejecta.  Stronger  
lines are optically thick through more of the ejecta, possibly
including the less dense spaces between clumps.  If
the $\lambda$5169 line (with the largest $gf$-value) is optically
thick over more of the ejecta than the other \ion{Fe}{2} lines, then
the signature of the asphericity  
could be hidden.  Another possibility is that blending with another
line (\ion{Si}{2} $\lambda$5051 at photospheric velocities is a possible
candidate) destroys the polarization signature.

\subsubsection{Non-detections}

It is also worth considering which line features do not show
spectropolarimetric features, as well as those that do.  The
S/N of our observations degrades rapidly blueward of
4000 \AA\ due to declining efficiency of the spectrograph and
increasing atmospheric absorption.  Despite these mitigating effects,
we are confident that Figure~\ref{spolfig} 
shows no sign of strong spectropolarimetric features associated with the
\ion{Ca}{2} H\&K lines of the sort seen in the NIR triplet lines of the
same species.  One potential explanation is that, if our choice
of ISP is correct, the continuum at such short wavelengths has
negligible polarization due to the high opacity in lines from
iron-peak elements.  The additional opacity of the 
strong H\&K lines then results in little additional depolarization.

Two of the more interesting ionic species that lack strong
polarimetric signatures are \ion{O}{1} and \ion{S}{2}.  Oxygen
in the ejecta of SNe~Ia could either be unprocessed material from the
progenitor white dwarf or the result of incomplete carbon burning.  In
either case the oxygen should not share the same spatial distribution
as the silicon.  SN~2004dt \citep{leo05,wa06} showed the combination
of strong  polarization in the HV \ion{Si}{2} line features and a lack
of a feature at the \ion{O}{1} line, which \citet{wa06} interpreted to
be a result of turbulent burning processes leaving clumpy intrusions
of silicon in an oxygen-dominated substrate.  The \ion{O}{1}
$\lambda$7773 line in \sn\ is of normal strength (see
Figure~\ref{compfig}), with a flux minimum near a radial velocity of
9,000 \kms.   No strong polarization feature is visible in
Figure~\ref{spolfig} in $u_{RSP}$, and there is marginal evidence for
an increase in $q_{RSP}$.  Unfortunately, this line falls near the
strongest atmospheric absorption at optical wavelengths (the A band)
and so the S/N is somewhat degraded.  

Sulfur, however, should have a spatial distribution similar to that of
silicon, as both elements are formed by similar burning processes.
\ion{S}{2} is responsible for the characteristic 
``W''-shaped feature near 5300 \AA\ in SNe~Ia before maximum light, as
well as more minor features just blueward of that.  At this late
epoch, the ``W''-shaped feature has started to fade and the redder
absorption is being dominated by the increased effect of iron on the
spectrum at these wavelengths.  We see no strong 
modulation in $q_{RSP}$ at the position of the \ion{S}{2}
features and $u_{RSP}$ definitely lacks the strong feature seen in the
\ion{Si}{2} line.  Differences in polarization behavior between the
two species despite similar spatial distributions could be a result of
different optical depths in the lines, or the underlying continuum
polarization could be modified differently by the blends of lines from
iron-peak elements.

\section{Discussion}

The most basic question to be answered by spectropolarimetry is
whether or not an object is spherically symmetric.  The strong
wavelength dependence of the polarimetric data shown above is
convincing evidence of intrinsic polarization and hence of deviations
from spherical symmetry in the ejecta of \sn.

With the set of Stokes parameters that we have chosen to present the
data, $q_{RSP}$ would be an estimator of the total polarization
if the system had a well-defined symmetry axis.
The need for a second variable to describe the data, here the Stokes
parameter $u_{RSP}$, is indicative of deviations from axisymmetry.
The second parameter is significantly non-zero only for wavelength
regions near certain spectral features.  Equivalently, the scatter of
the continuum data points about the best-fit line in
Figure~\ref{qufig} is similar to that 
expected from the errors.  This means the SN ejecta can
be approximated as having a global, axisymmetric asphericity with
deviations caused by a non-axisymmetric distribution of certain ionic
species.

Polarization in a SN atmosphere is induced by light scattering off of
electrons.  Thomson scattering is a wavelength-independent process,
but our estimator of the total polarization, $q_{RSP}$, shows a number
of broad undulations with wavelength and a general decrease in the
polarization level toward the blue end of the spectrum.  This is due
to the nature of the opacity in SNe~Ia, which is
dominated by a large number of bound-bound transitions, primarily of
iron-peak elements \citep{pe00b}.  These transitions act to depolarize
the emergent radiation \citep{ho96}, so the gradient of polarization
percentage from red to blue reflects the increasing number of such
depolarizing transitions in the blue and ultraviolet portion of the
spectrum \citep{ho01}.  Note that this 
interpretation does depend on our choice for the ISP, which we will
revisit at the end of this section.

The detailed form of the undulations seen in $q_{RSP}$ is therefore a
direct probe of the line opacity in the continuum-formation region of
SNe~Ia, as wavelengths of greater 
line opacity should show more depolarization.  Inspection of plots of
the expansion opacity given in \citet{pe00b} and \citet{ka06b} shows a
local maximum in the line opacity 
near 5300 \AA\ and a minimum near 4600 \AA, similar to the wavelengths
of local minima and maxima, respectively, in the polarization data
shown in Figure~\ref{spolfig}.  This correspondence may only be
suggestive, but can be verified by modern 3D radiative
transfer calculations of the type described by Kasen et al. (2006).

The amplitude of continuum polarization gives information about the
global degree of asphericity of the SN atmosphere
\citep{ss82,je91,ho91}.
SN~1999by was the first SN~Ia with a firm detection of intrinsic
polarization, at least 0.7\% \citep{ho01}.  The best-fit model of
\citet{ho01} explained the polarization as being due to an oblate
ellipsoid with the major axis 17\%  longer than the minor axis and
viewed nearly equator-on.  Despite the sometimes large line
polarizations, subsequent observations of SNe~Ia have yielded
detections of intrinsic continuum polarization that
are smaller than in SN~1999by, with values of 0.3--0.4\% being
typical \citep{wa03,wa06,leo05,me06}.  Converting the
continuum polarization to a geometry is model-dependent (e.g.,
H\"{o}flich 1991), but the
lower polarizations of most SNe~Ia may be due to objects with
asphericities closer to 10\% \citep{wa03}.  \sn\ is no exception, with
a minimum of 0.4\% intrinsic continuum polarization.

The polarization angle rotations described above in line features of
\ion{Si}{2}, \ion{Ca}{2}, and \ion{Fe}{2} are evidence that the
geometric distribution of these ionic species within the ejecta
differs from the geometry of the continuum-emitting SN atmosphere.
The pure 
ellipsoidal models of SN~1999by explored by \citet{ho01} had only
radial variations in composition, but they were able to successfully
model the spectropolarimetric feature associated with \ion{Si}{2}
present in their data.  The additional opacity present in the strong
line relative to nearby wavelengths resulted in additional
depolarization at a constant polarization angle.  Such a model
for \sn\ could 
possibly explain the dip in $q_{RSP}$ seen in our data centered on the
\ion{Si}{2} line, but not the feature in $u_{RSP}$.  In particular,
the loop in the $q-u$ plane for the feature associated with the
\ion{Ca}{2} NIR triplet cannot be formed in a simple axisymmetric model
for the ejecta.

The HV material in SN~2001el also showed such a $q-u$ loop
\citep{wa03, ka03}.  \citet{ka03} explored several different
geometries for the HV ejecta and found that $q-u$ loops were a generic
consequence of systems whose components had different axes of
symmetry.  For example, they found that an ellipsoidal shell whose
axis of symmetry 
was misaligned with respect to an ellipsoidal continuum region
produced such a loop.  They also found that models invoking clumps in
the ejecta could produce $q-u$ loops.  Clumps can produce polarimetric
signatures by selectively blocking and 
depolarizing part of the light from the underlying polarized
continuum.  These clumps might not be oriented along the symmetry axis
of the system, and hence produce polarimetric angle rotations.

In \sn, however, the main line-polarization features are being
formed at photospheric velocities.  This can be seen most clearly in
the \ion{Si}{2} panels of Figure~\ref{multifig}.  While modulations of
the polarization exist at all wavelengths, the wavelength bins with the
greatest polarization deviations from that of the continuum are
located just blueward of 
the P-Cygni absorption minimum.  The main source of polarization could
be clumps of \ion{Si}{2}-absorbing material selectively blocking
portions of the polarized photosphere and confined rather narrowly in
velocity space above the photospheric velocity.
Support for this picture comes from the two separate polarimetric
features seen in the \ion{Ca}{2} NIR triplet.  Explanations for the
line polarization must prevent the polarization feature 
associated with $\lambda$8662 from blending with the polarization
feature formed by $\lambda\lambda$8498, 8542, just 4000 \kms\ to the
blue.  Depolarizing clumps in the ejecta confined to velocities less
than $\sim$13,000 \kms\ could be responsible.

\citet{leo05} described spectropolarimetric angle rotations in line
features of \ion{Si}{2} and \ion{Ca}{2} in SN~2003du that are at least
broadly similar to 
these results.  They ascribed the rotations to either clumps in the
ejecta or ionization asymmetries of the sort invoked by \citet{ch92}
to explain observations of SN~1987A.  Clumps of radioactive $^{56}$Ni
formed in the explosion could produce local variations in ionization
and electron density that would be manifested as polarization
features.  In addition, \citet{wa03} describe a potential ``skin
effect'' as $^{56}$Ni clumps produce a tangential component to the
radiation field whose effects would vary as a function
of wavelength due to the wavelength dependence of the photospheric
radius.  However, any clumps of $^{56}$Ni would only have a modest
effect on the ionization state of SN~Ia ejecta and
the significant path length of gamma rays in the ejecta would tend to
smooth out the effects of clumpy energy deposition from radioactive
elements.  Therefore, we do not favor the interpretation where
polarimetric effects are directly due to the clumpy distribution of
$^{56}$Ni instead of the clumpy distribution of those ions directly
responsible for the polarimetric features.

Clumps of intermediate-mass elements could be detectable by other
means.  \citet{th02} examined the consequences of clumping on
variation in line profiles.  In their parameterized models, the
observed homogeneity of \ion{Si}{2} line profiles in SNe~Ia meant that
the characteristic size scale of perturbations in composition must be
of order 1\% of the size of photosphere or smaller.  Larger
perturbations would result in variations of the line profile when
observed from different orientations that are larger than the observed
variation in normal SNe~Ia, though we note that in this particular
object the \ion{Si}{2} line is somewhat 
weaker than normal (\S\ref{fluxsect}; Krisciunas et al. 2007).  A
single clump in front of  
the SN photosphere can result in large observed polarizations
\citep{ka03}, so perhaps the small value of the polarization
modulations in the lines is a result of many small clumps, consistent
with the \citet{th02} results.  Future numerical work will make
possible quantitative calculations of the effects of clumps of
different size scales on the polarization spectrum \citep{ka06a}.

Next we turn our attention to the HV lines.  The evidence presented
above shows that the line-polarization features in \sn\ are associated
with material near the photospheric velocity, not the HV line-forming
region.  A small polarization modulation is visible in $u_{RSP}$ in
the blue wing of the HV \ion{Ca}{2} NIR triplet feature.  A formal
integration over 7875--8075 \AA\ (blueshifts of 17,000--24,000
\kms\ relative to $\lambda$8542) gives a non-zero measurement of
$u_{RSP}=0.103 \pm\ 0.025 \%$, which is significantly smaller than the
photospheric feature.  Naively,
this is somewhat unexpected as SNe 2001el, 2002bf, and 
2004dt all showed strong (0.7\% $< P <$ 2.5\%) polarization features in
their HV features \citep{wa03,leo05,wa06}.  The answer may be that we
caught \sn\ at a special epoch, when the HV material had sufficient
optical depth (or a large enough photospheric covering fraction) to
imprint an absorption feature on the spectrum, but not enough to result 
in a significant polarization feature.

The time series of spectropolarimetry
presented in \citet{wa03} showed that the HV polarization feature in
SN 2001el was strong at early times and disappeared after maximum
light as the feature in the flux spectrum also faded.  Our data are
most closely comparable in time to their 2001 October 9 epoch (+10~d 
after maximum light), in which the HV flux feature was still
present, but fading, and the polarization feature was 
mostly gone.  SN~2003du provides another example of this phenomenon.
\citet{ger04} presented spectra at early times that showed a strong HV
feature, but the data analyzed by \citet{leo05}, taken
18~d after maximum light, show no HV absorption or associated
polarization feature.

A comparison between the polarimetry of SNe~2001el and 2004S is very
interesting.  \citet{leo05} found that the spectropolarimetric
properties of SNe~Ia were diverse, but may be correlated with the
photometric and spectroscopic 
properties.  In this case, \citet{kk06} demonstrated that these two
supernovae have almost identical light curves and spectra, so we would
like to know whether  
the similarities extend to the polarimetry as well.  Fortuitously, our
single set of spectropolarimetry is at almost the same epoch (+9~d) 
as one of the observations of SN~2001el presented by
\citet{wa03} (+10~d), though we do not have a corresponding time
series of data.  \citet{wa03} used a different method to determine the
ISP (see below) so our results are not directly comparable, but we
can make qualitative comparisons to the data presented in their
Figures 9, 10, and 11.

The continuum polarization of SN~2001el has a maximum amplitude of
0.3\% and it decreases rather sharply by a few tenths of a percent
blueward of $\sim$5400 \AA, which is where \sn\ also shows a
change in continuum polarization.  At earlier epochs, 
the \ion{Si}{2} $\lambda$6355 line in SN~2001el shows deviations from
axisymmetry in the form of a $q-u$ loop with an amplitude of 0.5\%,
similar to what we see in our \sn\ data, though the feature in
SN~2001el mostly faded by +10~d to the $\sim$0.1\% level.  Some
difference in the silicon distribution between the two SNe should be
expected from the slightly weaker and redder \ion{Si}{2} line in the
flux spectrum of \sn\ \citep{kk06}.

As noted above, the HV \ion{Ca}{2} NIR triplet polarization feature in
SN~2001el largely disappeared between maximum light and +10~d.  The
photospheric component of the NIR triplet in SN~2001el has a few
tenths of a percent polarization at +10~d, although it lacks the
two separate features seen in \sn.  SN~2001el also shows no obvious
sign of the $q-u$ loop at photospheric velocities apparent in the
\ion{Ca}{2} feature of \sn.  A 
polarization feature (0.2\% in each Stokes parameter) in SN~2001el
near 4800 \AA\ on day +10 is not mentioned by \citet{wa03}, but
appears to correspond to the features in \sn\ which we have associated
with \ion{Fe}{2}.  In summary,
SN~2001el has slightly less continuum polarization than \sn\ and weaker
line-polarization features at the epoch most comparable to our data,
though the \ion{Si}{2} line in SN~2001el at earlier epochs is similar
to our data for \sn.

The interpretation of the results presented in this paper is affected
by uncertainty in the determination of the ISP.  Our choice of ISP has
been guided by the belief that the intrinsic polarization at
wavelengths less than 5000 \AA\ should be minimal, as detailed by 
\citet{ho01}.  Another plausible method for estimating the ISP is to
obtain observations at a very late epoch, when the electron scattering
optical depth should be negligible and any measured polarization can be
ascribed to ISP.  \citet{wa03} obtained such observations for
SN~2001el and determined a different value for the ISP than was measured
from the first method \citep{ka03}, with no obvious way to reconcile
the two possibilities.  We lack late-time observations of \sn, so we
cannot perform this test.

In this work we have chosen the first ISP method due to the simplicity
of interpretation and the lack of a compelling argument for another
ISP choice, but for the rest of this section we will examine the
consequences of relaxing our assumptions.  If we do not choose an ISP
along the main axis shown by the continuum data points in
Figure~\ref{qufig}, then the angle of polarization of the continuum
will vary with wavelength.  One possible alternative ISP would be near
the point ($q_{RSP}$,$u_{RSP}$) of (0.0,$-0.3$) in
Figure~\ref{linepolfig}, chosen to minimize the polarization angle
rotation through the \ion{Si}{2} line as well as the magnitude of
the polarization at the minimum of the P-Cygni absorption, but this
ISP vector creates problems of interpretation as well.
No choice for the ISP can eliminate the $q-u$ loop shown by the
\ion{Ca}{2} NIR triplet.  If we know polarization angle rotations must
occur across a feature due to one intermediate-mass element, then
perhaps we should expect them in others as well.  In addition, the
total host-galaxy ISP correction would be uncomfortably large given 
the low host-galaxy reddening.

Potential ISP vectors other than our chosen one necessarily lead to
non-zero polarization at wavelengths below 4000 \AA, which would be
an important result.  Implicit in our discussion of the continuum
polarization is a construction whereby continuum radiation is polarized
by electron scattering and then subsequently depolarized by bound-bound
line-transport processes in the outer part of the SN atmosphere.  This
picture is backed by detailed numerical simulations of radiation
transport in SNe~Ia.  \citet{pa96} analyzed the 1D
deflagration model W7 \citep{nom84} and found that at maximum light
the optical depth to electron scattering was of order unity near a
velocity of 5000 \kms.  The apparent ``photospheric'' velocities
measured from the minima of P-Cygni line features are closer to 10,000
\kms\ due the effects of pseudo-continuous opacity from many weak
lines of iron-peak elements.

At the later epoch of our \sn\ observations (+9~d), 
the surface of optical depth unity to electron scattering
should have receded even more deeply into the ejecta.  In reality, the
competition between polarizing electron scattering 
and depolarizing lines occurs over an extended range of
radii.  The predominance of line opacity over Thomson 
scattering \citep{pe00b} means that at wavelengths of high line
opacity, the last scattering of photons occurs at larger
radii in the ejecta (e.g., H\"{o}flich et al. 1998),
which limits the opportunity for those photons to be subsequently
scattered by electrons and polarized as they exit the SN
atmosphere.

These arguments are based on 1D properties of SN~Ia
simulations.  Non-zero polarization at wavelengths of high line
opacity could be a 
sign of clumps of $^{56}$Ni mixed out into the ejecta.  Such clumps
could result in local enhancements of ionization and hence electrons
that could produce polarization at larger radii than expected.
Significant outward mixing of $^{56}$Ni would be an interesting
constraint on explosion models and would also have effects on the
photometric rise time and overall light-curve shape.  Choosing an ISP
vector away from the line in Figure~\ref{qufig} also
produces a slow rotation of the continuum polarization angle with
wavelength that is hard to understand, though \citet{wa03} posit 
a potential ``skin effect''.    The
time dependence of the polarization 
effects due to these hypothetical $^{56}$Ni clumps is unclear, though
obviously our single epoch of observations cannot provide any
constraints by itself.

\section{Conclusions}

We have presented a single epoch of high S/N
spectropolarimetry of the Type Ia \sn, which had an intrinsic
continuum polarization of 0.4\%.  This observation adds to a growing
body of evidence that normal SNe~Ia have global asphericities of order
10--20\%.  The data points from the continuum are consistent with an
axisymmetric geometry such as an ellipsoid.  Polarization angle
modulations across spectral features due to \ion{Si}{2}, 
\ion{Ca}{2}, and \ion{Fe}{2} are evidence that the distributions of
some elements within the supernova ejecta differ from that of
the continuum.  One natural interpretation of these data is that we
are seeing evidence for clumps of newly synthesized elements.  These
clumps could be a diagnostic of turbulent burning during the explosion
process (e.g., Gamezo et al. 2005).  Previous work has shown
non-axisymmetric structures in HV material \citep{ka03}, but here the
deviations from axisymmetry are present close to the photosphere.

We argued that the clumps in the ejecta are sufficiently narrowly
confined in velocity space to prevent the \ion{Ca}{2} NIR triplet
polarization features from being blended together.  This conclusion
could be tested with a time series of spectropolarimetric observations
of a future SN~Ia starting well before maximum light.  At early
times these clumps might be hidden because the lines are optically
thick at higher velocities.  As the absorption minima
in the flux spectra recede through the ejecta, the 
clumps could become visible polarimetrically at later times.  In
particular, the spectroscopic sequence of \sn\ presented by
\citet{kk06} shows that at early times the flux minimum of the
\ion{Ca}{2} NIR triplet line was at higher velocity than the
polarimetric features we see at this later epoch.  Spectropolarimetric
data of the similar SN~2001el before maximum light \citep{wa03} show
no features at photospheric velocities in the \ion{Ca}{2} NIR triplet
(which has a strong HV component), even though photospheric
polarimetric features are visible in the \ion{Si}{2} $\lambda$6355
line (which does not).

A time series of spectropolarimetric data would also be useful in
understanding the HV material.  \sn\ is the first SN~Ia observed
polarimetrically with prominent HV features in the total-flux spectrum 
that lacked counterparts in the polarization.  We argued above that this
was likely a consequence of the timing of our observations, as the HV
polarization features in the similar SN~2001el also disappeared a
couple of weeks after maximum light \citep{wa03}.  The HV material
must decrease in optical depth as it expands, so determining the epoch
at which its effects disappear in both the total-flux
spectrum and the polarization data has the potential to constrain
the optical depth or covering fraction of the HV material, key
parameters in any model.  Alternatively, we could be viewing the HV
material from a special orientation, such as along an axis of
symmetry.

The deviations from axisymmetry in \sn\ are responsible for subtle 
variations in the polarization across the profiles of some spectral
lines and would have been unobservable at lower S/N.
Spectropolarimetric observations have the potential to be a 
powerful probe of SN~Ia explosion models; however, quantitative
conclusions will have to wait for detailed radiative transfer
calculations of current 3D explosion models (e.g., Kasen et
al. 2006).  In addition, interpretation of the continuum observations
is hampered by uncertainties in the removal of ISP, even for objects
like \sn\ suffering small extinction.  Progress can be made with
observations of a larger set of objects, preferably behind minimal
dust columns. 

\acknowledgments
We are grateful to R.~J. Foley for assistance with the
observations and K. Krisciunas for sharing results in advance of
publication.  We also would like to thank the referee, D. Kasen, for
his close reading of this manuscript.
The data presented herein were obtained at the W.~M. Keck Observatory,
which is operated as a scientific partnership among the California
Institute of Technology, the University of California and the National
Aeronautics and Space Administration. The Observatory was made
possible by the generous financial support of the W. M. Keck
Foundation.  We wish to recognize and acknowledge the very
significant cultural role and reverence that the summit of Mauna Kea
has always had within the indigenous Hawaiian community; we are most
fortunate to have the opportunity to conduct observations from this
mountain.  We also would like to thank the expert assistance of the
Keck staff in making these observations possible. This research
was supported by National Science Foundation grants AST--0307894 
and AST--0607485.

\clearpage

\clearpage

\clearpage

\clearpage

\clearpage

\clearpage

\begin{thebibliography}{}

\bibitem[Benetti et al.(2005)]{ben05}Benetti, S., et al. 2005, ApJ, 623,
1011 

\bibitem[Berdyugin et al.(1995)]{berd95}Berdyugin, A.,
  Sn\aa re, M.-O., \& Teerikorpi, P. 1995, \aap, 294, 568

\bibitem[Branch et al.(1988)]{br88}Branch, D., Drucker, W., \&
  Jeffery, D.~J. 1988, ApJ, 330, L117

\bibitem[Branch et al.(1995)]{br95}Branch, D., Livio, M., Yungelson,
  L.~R., Boffi, F.~R., \& Baron, E. 1995, PASP, 107, 1019

\bibitem[Branch et al.(2006)]{br06}Branch, D., et al. 2006, PASP,
  118, 560

\bibitem[Chornock et al.(2006)]{me06}Chornock, R., Filippenko, A.~V.,
  Branch, D., Foley, R.~J., Jha, S., \& Li, W. 2006, PASP, 118, 722

\bibitem[Chugai(1992)]{ch92}Chugai, N.~N. 1992, Soviet Astron. Lett.,
  18, 168

\bibitem[Clemens \& Tapia(1990)]{ct90}Clemens, D.~P., \& Tapia,
  S. 1990, PASP, 102, 179

\bibitem[Filippenko(1982)]{fi82}Filippenko, A.~V. 1982, PASP, 94, 715

\bibitem[Filippenko(1997)]{fil97}Filippenko, A.~V. 1997, ARAA, 35, 309 

\bibitem[Gamezo et al.(2005)]{gam05}Gamezo, V.~N.,
  Khokhlov, A.~M., \& Oran, E.~S.  2005, ApJ, 623, 337

\bibitem[Gamezo et al.(2003)]{gam03}Gamezo, V.~N., Khokhlov, A.~M.,
  Oran, E.~S., Chtchelkanova, A.~Y., \& Rosenberg, R.~O. 2003,
  Science, 299, 77

\bibitem[Gerardy et al.(2004)]{ger04}Gerardy, C., et al. 2004, ApJ,
  607, 391

\bibitem[Hatano et al.(1999)]{ha99}Hatano, K., Branch, D., Fisher, A.,
  Baron, E., \& Filippenko, A.~V. 1999, ApJ, 525, 881

\bibitem[Hillebrandt \& Niemeyer(2000)]{hn00}Hillebrandt, W., \&
  Niemeyer, J.~C. 2000, ARAA, 38, 191

\bibitem[H\"{o}flich(1991)]{ho91}H\"{o}flich, P. 1991, \aap, 246, 481

\bibitem[H\"{o}flich et al.(1996)]{ho96}H\"{o}flich, P., Wheeler,
  J.~C., Hines, D.~C., \& Trammell, S.~R.  1996, ApJ, 459, 307

\bibitem[H\"{o}flich et al.(1998)]{ho98}H\"{o}flich, P., Wheeler,
  J.~C., \& Thielemann, F.~K. 1998, ApJ, 495, 617

\bibitem[Howell et al.(2001)]{ho01}Howell, D.~A., H\"{o}flich, P.,
  Wang, L., \& Wheeler, J.~C. 2001, ApJ, 556, 302

\bibitem[Jeffery(1989)]{jef89}Jeffery, D.~J. 1989, ApJS, 71, 951

\bibitem[Jeffery(1991)]{je91}Jeffery, D.~J. 1991, ApJ, 375, 264

\bibitem[Kasen(2006)]{ka06b}Kasen, D. 2006, ApJ, 649, 939

\bibitem[Kasen \& Plewa(2005)]{kp05}Kasen, D., \& Plewa, T.  2005, ApJ,
  622, L41

\bibitem[Kasen et al.(2006)]{ka06a}Kasen, D., Thomas, R.~C., \&
  Nugent, P. 2006, ApJ, 651, 366

\bibitem[Kasen et al.(2003)]{ka03}Kasen, D., et al. 2003, ApJ, 593,
  788

\bibitem[Khokhlov(1991)]{kh91}Khokhlov, A.~M. 1991, \aap, 245, 114

\bibitem[Kozma et al.(2005)]{koz05}Kozma, C., Fransson, C.,
  Hillebrandt, W., Travaglio, C., Sollerman, J., Reinecke, M.,
  R\"{o}pke, F.~K., \& Spyromilio, J. 2005, \aap, 437, 983

\bibitem[Krisciunas et al.(2007)]{kk06}Krisciunas, K., et al. 2007,
  AJ, 133, 58

\bibitem[Leibundgut et al.(1993)]{lei93}Leibundgut, B., et al. 1993,
AJ, 105, 301

\bibitem[Leonard et al.(2001)]{le01}Leonard, D.~C., Filippenko, A.~V.,
  Ardila, D.~R., \& Brotherton, M.~S. 2001, ApJ, 553, 861

\bibitem[Leonard et al.(2005)]{leo05}Leonard, D.~C., Li, W.,
  Filippenko, A.~V., Foley, R.~J., \& Chornock, R. 2005, AJ, 632, 450 

\bibitem[Li et al.(2003)]{li03}Li, W., et al. 2003, PASP, 115, 453

\bibitem[Livio(2001)]{liv01}Livio, M. 2001, in Supernovae and
  Gamma-ray Bursts: The Greatest Explosions Since the Big Bang,
  ed. M. Livio, N. Panagia, \& K. Sahu (Cambridge: Cambridge
  Univ. Press), 334

\bibitem[Martin \& Biggs(2004)]{mb04}Martin, R., \& Biggs, J. 2004,
  \iaucirc, 8282, 1

\bibitem[Matheson et al.(2000)]{ma00}Matheson, T., Filippenko, A.~V.,
  Ho, L.~C., Barth, A.~J., \& Leonard, D.~C. 2000, AJ, 120, 1499

\bibitem[Mathewson \& Ford(1970)]{mat70}Mathewson, D.~S., \& Ford,
  V.~L. 1970, \memras, 74, 139

\bibitem[Mazzali et al.(2005a)]{maz05a}Mazzali, P., Benetti, S.,
  Stehle, M., Branch, D., Deng, J., Maeda, K., Nomoto, K., \& Hamuy,
  M. 2005a, MNRAS, 357, 200.

\bibitem[Mazzali et al.(2005b)]{maz05b}Mazzali, P., et al. 2005b, ApJ,
  623, L37

\bibitem[Miller et al.(1988)]{mrg88}Miller, J.~S., Robinson, L.~B., \&
  Goodrich, R.~W. 1988, in Instrumentation for Ground-Based Astronomy,
  ed. L.~B. Robinson (New York: Springer-Verlag), 157

\bibitem[Misra et al.(2005)]{mi05}Misra, K., Kamble, A.~P.,
  Bhattacharya, D., \& Sagar, R. 2005, MNRAS, 360, 662

\bibitem[Nomoto et al.(1984)]{nom84}Nomoto, K., Thielemann, F.~K., \&
  Yokoi, K. 1984, ApJ, 286, 644

\bibitem[Nomoto et al.(2003)]{nom03}Nomoto, K., Uenishi, T.,
  Kobayashi, C., Umeda, H., Ohkubo, T., Hachisu, I., \& Kato, M.
  2003, in From Twilight to Highlight: The Physics of Supernovae,
  ed. W. Hillebrandt \& B. Leibundgut (Berlin:Springer-Verlag), 115

\bibitem[Oke et al.(1995)]{oke95}Oke, J.~B., et al. 1995, PASP, 107,
  375

\bibitem[Pauldrach et al.(1996)]{pa96}Pauldrach, A.~W.~A., Duschinger,
  M., Mazzali, P.~A., Puls, J., Lennon, M., \& Miller, D.~L. 1996,
  \aap, 312, 525

\bibitem[Phillips(1993)]{ph93}Phillips, M.~M. 1993, ApJ, 413, L105

\bibitem[Pinto \& Eastman(2000)]{pe00b}Pinto, P.~A., \& Eastman,
  R.~G. 2000, ApJ, 530, 757

\bibitem[Reinecke et al.(2002)]{rhn02}Reinecke, M.,
  Hillebrandt, W., \& Niemeyer, J.~C. 2002, \aap, 391, 1167

\bibitem[Schlegel et al.(1998)]{sfd98}Schlegel, D.~J.,
  Finkbeiner, D.~P., \& Davis, M. 1998, ApJ, 500, 525

\bibitem[Schmidt et al.(1992)]{sch92}Schmidt, G.~D., Elston, R., \&
  Lupie, O.~L. 1992, AJ, 104, 1563

\bibitem[Serkowski et al.(1975)]{ser75}Serkowski, K., Mathewson,
  D.~S., \& Ford, V.~L. 1975, ApJ, 196, 261

\bibitem[Shapiro \& Sutherland(1982)]{ss82}Shapiro, P.~R., \& Sutherland,
  P.~G. 1982, ApJ, 263, 902

\bibitem[Stehle et al.(2005)]{st05}Stehle, M., Mazzali, P.~A.,
  Benetti, S., \& Hillebrandt, W. 2005, MNRAS, 360, 1231

\bibitem[Suntzeff et al.(2004)]{su04}Suntzeff, N., Globus, A., Galli,
  L., Whiting, A., \& Schmidtobreick, L. 2004, \iaucirc, 8283, 1

\bibitem[Theureau et al.(2005)]{th05}Theureau, G., et al. 2005, \aap,
  430, 373

\bibitem[Thomas et al.(2002)]{th02}Thomas, R.~C., Kasen, D., Branch,
  D., \& Baron, E. 2002, ApJ, 567, 1037

\bibitem[Trammell et al.(1993)]{tdg93}Trammell,
  S.~R., Dinerstein, H.~L., \& Goodrich, R.~W. 1993, ApJ, 402, 249

\bibitem[Tran(1995)]{tr95}Tran, H.~D. 1995, ApJ, 440, 565

\bibitem[Wang et al.(2006)]{wa06}Wang, L., Baade, D., H\"{o}flich, P.,
  Wheeler, J.~C., Kawabata, K., Khokhlov, A., Nomoto, K., \& Patat,
  F. 2006, ApJ, 653, 490

\bibitem[Wang, Baade, \& Patat(2007)]{wa07}Wang, L., Baade, D., \& Patat, F.
  2007, Science, 315, 212
  
\bibitem[Wang et al.(2001)]{wa01}Wang, L., Howell, D.~A., H\"{o}flich,
  P., \& Wheeler, J.~C. 2001, ApJ, 550, 1030

\bibitem[Wang et al.(2003)]{wa03}Wang, L., et al. 2003, ApJ, 591, 1110

\bibitem[Wells et al.(1994)]{we94}Wells, L.~A., et al. 1994, AJ, 108,
  2233

\end{thebibliography}
\end{document}